\documentclass[12pt]{article}
\usepackage{graphicx}

\newcommand{\beq}{\begin{equation}}
\newcommand{\eeq}{\end{equation}}
\newcommand{\beqn}{\begin{eqnarray}}
\newcommand{\eeqn}{\end{eqnarray}}
\newcommand{\beqns}{\begin{eqnarray*}}
\newcommand{\eeqns}{\end{eqnarray*}}

\begin{document}

\begin{titlepage}
\begin{center}

\hfill USTC-ICTS/PCFT-23-04\\
\hfill January 2023

\vspace{2.5cm}

{\large {\bf Rare $W$-boson decays into a vector meson and lepton pair}}\\
\vspace*{1.0cm}
 {Dao-Neng Gao$^\dagger$\vspace*{0.3cm} \\
{\it\small Interdisciplinary Center for Theoretical Study,
University of Science and Technology of China, Hefei, Anhui 230026
China}\\
{\it\small Peng Huanwu Center for Fundamental Theory, Hefei, Anhui 230026 China}}

\vspace*{1cm}
\end{center}
\begin{abstract}
\noindent
We have presented a theoretical study of exclusive rare $W$-boson decays, $W\to V\ell\bar{\nu}_\ell$ with $V$ denoting a neutral vector meson and $\ell=e$ or $\mu$, in the standard model. The leading-order contributions to these processes are given by $W\to \gamma^*\ell\bar{\nu}_\ell$ with the subsequent $\gamma^*\to V$ transition.
 Branching fractions of these decay modes, for $V=\rho$, $\omega$, $\phi$, and $J/\Psi$, respectively, have been calculated and predicted around $10^{-6}\sim 10^{-7}$, which are surprisingly larger than those of two-body hadronic radiative decays $W^\pm\to M^\pm \gamma$ with $M$ denoting a pseudoscalar or vector meson. Thus it is expected that rare $W$ decays into a neutral vector meson plus lepton pair may be the promising channels in future experimental facilities with a large number of $W$-boson events produced.

\end{abstract}

\vfill
\noindent
$^{\dagger}$ E-mail address:~gaodn@ustc.edu.cn
\end{titlepage}
\vspace{0.5cm}

%\section{Introduction}
Exclusive rare $W$-boson decays, which contain hadronic final states, could provide interesting probes to increase our understanding of the properties of the fundamental weak gauge boson as well as offer some deep insights into quantum chromodynamics \cite{AMP82,MM15,GKN15, KP94, IJXY19}. Experimentally, no such processes have been observed so far, and only upper limits on the branching fractions of three exclusive modes: ${\cal B}(W^\pm\to D_s^\pm\gamma)<1.3\times 10^{-3}$, ${\cal B}(W^\pm\to \pi^\pm\gamma)<7\times 10^{-6}$, and ${\cal B}(W^\pm\to \pi^+\pi^-\pi^\pm)<1.01\times 10^{-6}$, were set at 95\% confidence level \cite{PDG22}. On the other hand, a huge number of $W$ events, about ${\cal O}(10^{11})$, will be expectedly accumulated in the high-luminosity Large Hadron Collider (LHC) \cite{GKN15}. This may significantly facilitate the experimental studies of rare $W$-boson decay channels, which can be very helpful both to test the standard model (SM) and to search for new physics beyond the SM.

Our main focus in the present paper is on another types of rare $W$-boson decays: $W\to V \ell\bar{\nu}_\ell$ with $V$ denoting the neutral vector particle including heavy quarkonium $J/\Psi$ or light mesons $\rho$, $\omega$, and $\phi$ etc. $\ell$ is the lepton with $\ell=e$ or $\mu$. The leading-order Feynman diagrams contributing to these processes in the SM have been shown in Figure 1, in which the transitions can proceed through $W\to \gamma^*\ell\bar{\nu}_\ell$, followed by $\gamma^*\to \bar{q}q \to V$. This is similar to the case of $Z\to V \ell^+\ell^+$ decays, which have been studied in Refs. \cite{BR90, Fleming93-94, CMS18}.

 First let us go into the decay amplitude of $W^-\to V\ell^-\bar{\nu}_\ell$. Using the standard vertices $W\ell\bar{\nu}_\ell$, $\gamma q\bar{q}$, and $\gamma W W$, one can carry out the direct calculation for Figure 1, which gives
\beqn\label{amp1}
{\cal M}(W^-\to V \ell^-\bar{\nu}_\ell)&=& -\frac{e^2 g Q_V f_V}{2\sqrt{2}m_V}\epsilon_\mu(p)\epsilon_\nu^*(q)\bar{u}(k_1)\left[\frac{2 k_1^\nu\gamma^\mu+\gamma^\nu q\!\!\!/\gamma^\mu}{q^2+2k_1\cdot q}\right.\nonumber\\
&& \left. -\frac{(2k+q)^\nu\gamma^\mu+2 q^\mu \gamma^\nu-2q\!\!\!/g^{\mu\nu}}{q^2+2 k\cdot q}\right](1-\gamma^5)v(k_2),
\eeqn
where $p$, $q$, $k_1$, and $k_2$ represent the momenta of $W^-$ and the final particles including $V$, $\ell^-$, and $\bar{\nu}_\ell$, respectively. $k=k_1+k_2$ denotes the momentum sum of lepton pair. $e$ is the QED coupling constant and $g$ is the weak SU(2)$_L$ coupling constant. $f_V$ is the decay constant of the vector meson, which is defined by
\beq\label{fv}
\langle V(p,\epsilon)|\bar{q}\gamma_\nu q |0\rangle= f_{V}m_{V}\epsilon^{*}_\nu.
\eeq
Here $\epsilon^{*}_\nu$ is polarization vector of $V$, and the value of $f_V$ can be extracted from the measured $V\to e^+e^-$ width. As shown in Ref. \cite{GKN15}, it has been already given that, $f_\rho=216.3\pm 1.3$ MeV, $f_\omega=194.2\pm 2.1$ MeV, $f_\phi=223.0\pm 1.4$ MeV, and $f_{J/\Psi}=403.3\pm 5.1$ MeV.  $Q_V$ is the quantity related to the electric charge of the quark inside $V$ with $Q_\rho=1/\sqrt{2}$, $Q_\omega=1/3\sqrt{2}$, $Q_\phi=-1/3$, and $Q_{J/\Psi}=2/3$.  Note that the use of the relation (\ref{fv}) in deriving eq. (\ref{amp1}) also fulfills the hadronization of the electromagnetic current $\bar{q}\gamma^\nu q$ into the final state particle $V$.

\begin{figure}[t]
\begin{center}
\includegraphics[width=12cm,height=4cm]{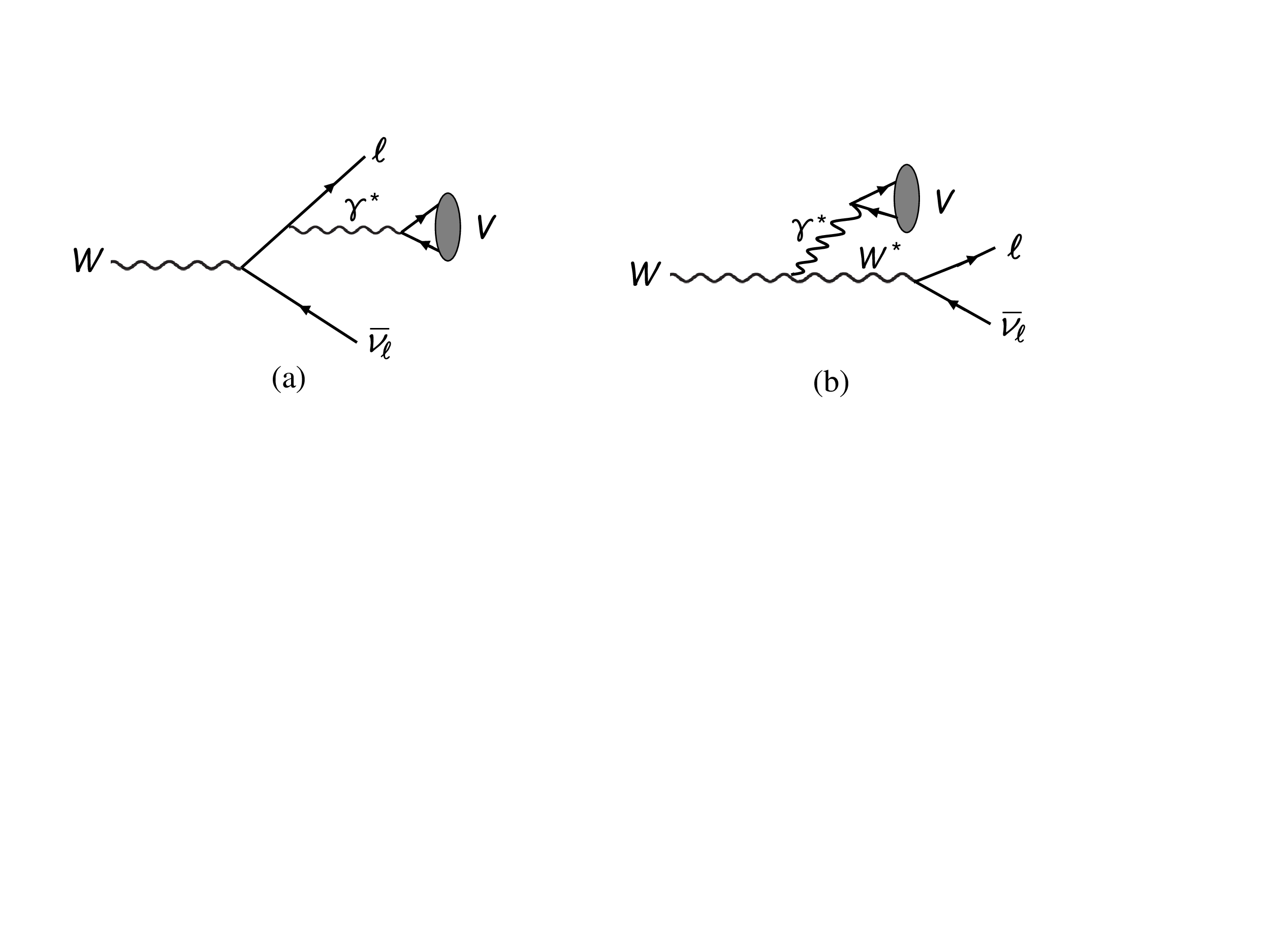}
\end{center}
\caption{The lowest-order Feynman diagrams for $W\to V \ell \bar{\nu}_\ell$ decays.}\label{figure1}
\end{figure}

Next, by squaring the decay amplitude (\ref{amp1}), summing or averaging the polarizations of final or initial particles, the differential decay rate of $W^-\to V\ell^-\bar{\nu}_\ell$ can be expressed as
\beqn\label{rate0}
\frac{d\Gamma}{d s_V d s_\ell}=\frac{m_W}{256\pi^3}\frac{1}{3}\sum_{\rm spins}|{\cal M}(W^-\to V \ell^-\bar{\nu}_\ell)|^2.\eeqn
Consequently, we get
\beq\label{rate1}
\frac{d\Gamma}{d s_V d s_\ell}=\frac{\alpha_{\rm em}^2 Q_V^2 g^2 f_V^2}{384\pi m_W r_V^2}I_V,
\eeq
where $\alpha_{\rm em}=e^2/4\pi$, $r_V=m_V/m_W$, and the lepton mass has been neglected in the calculation. The explicit expression of the dimensionless quantity $I_V$ is a little tedious, which will be shown in the Appendix. The Lorentz invariant dimensionless kinematical variables are defined as
\beq\label{kinematical-variable}
s_V\equiv(p-q)^2/m_W^2,\;\;\;\;\;s_\ell\equiv(p-k_1)^2/m_W^2,
\eeq
and the phase space can be given by
\beq\label{phasespace}
0\leq s_V\leq (1-s_\ell)(1-r_V^2/s_\ell),\;\;\;\;\; r_V^2\leq s_\ell \leq 1.
\eeq

Meanwhile, it is easy to compute the leading-order contribution to the width of pure leptonic $W$-boson decay for $\ell=e$ or $\mu$, which reads
\beq\label{leptonic decay}
\Gamma(W^-\to \ell^-\bar{\nu}_\ell)=\frac{g^2m_W}{48\pi}=\frac{G_F m_W^3}{6\sqrt{2}\pi}\equiv\Gamma_0,
\eeq
where $G_F$ is the Fermi constant given by $G_F/\sqrt{2}=g^2/8m_W^2$. Then one can choose to normalize the decay rate of $W^-\to V\ell^-\ell\bar{\nu}_\ell$ to $\Gamma_0$, which leads to
\beq\label{rate2}
\frac{1}{\Gamma_0}\frac{d\Gamma}{d s_V d s_\ell}=\frac{\alpha_{\rm em}^2 Q_V^2 f_V^2}{8 m_V^2}I_V.
\eeq
By further defining
\beq\label{YV}
Y_V\equiv\int I_V d s_V d s_\ell
\eeq
with the integral bound is given in eq. (\ref{phasespace}), one can get
\beq\label{width}
\frac{\Gamma(W^-\to V\ell^-\bar{\nu}_\ell)}{\Gamma_0}=\frac{\alpha_{\rm em}^2 Q_V^2 f_V^2}{8 m_V^2}Y_V.
\eeq

As mentioned above, the decay constants ($f_V$) of the neutral vector mesons have been extracted by the authors of Ref. \cite{GKN15} from the experimental data, and
\beq\label{vee}
\Gamma(V\to e^+ e^-)=\frac{4\pi Q_V^2 f_V^2}{3m_V}\alpha^2_{\rm em}(m_V)
\eeq
has been used. Therefore, after integrating over $I_V$ in eq. (\ref{YV}) to get $Y_V$, one can easily predict the decay rates of $W\to V \ell\bar{\nu}_\ell$ for $V=\rho$, $\omega$, $\phi$, and $J/\Psi$, respectively.

\begin{table}[t]\begin{center}\begin{tabular}{ c c c  c  c} \hline\hline
 $V$ & $m_{V}$(GeV)& $\Gamma(V\to e^+e^-)$(keV) & $Y_V$&$\Gamma(W^-\to V \ell^-\bar{\nu}_\ell)/\Gamma_0$ \\\hline
 $\rho$ & 0.775&$7.04\pm 0.06$ &$194.91$ & $(5.28\pm 0.04)\times 10^{-5}$\\
$\omega$& 0.782&$0.60\pm 0.02$ &$193.94$ &$(4.44\pm 0.15)\times 10^{-6}$\\
$\phi$& 1.019&  $1.27\pm 0.04$& $166.32$ & $(6.18\pm 0.19)\times 10^{-6}$\\
$J/\Psi$&3.097& $5.53\pm 0.10$&$74.53$&$(3.97\pm 0.07)\times 10^{-6}$\\\hline
\hline
\end{tabular}\caption{Decay rates of $W^-\to V \ell^-\bar{\nu}_\ell$ normalized to $\Gamma(W^-\to \ell^-\bar{\nu}_\ell)$ for $\ell=e$ or $\mu$. The values of $\Gamma(V\to e^+e^-)$ are taken from Ref. \cite{PDG22}.} \end{center}\end{table}

On the other hand, note that the scale of the electromagnetic coupling $\alpha_{\rm em}$ in eq. (\ref{rate2}) should also be at $m_V$ since, in Figure 1, this electromagnetic transition is via $\gamma^*\to V$. Therefore, combining eq. (\ref{width}) with eq. (\ref{vee}), one will obtain
\beq\label{rate3}
\frac{\Gamma(W^-\to V\ell^-\bar{\nu}_\ell)}{\Gamma_0}=\frac{3Y_V}{32\pi m_V}\Gamma(V\to e^+e^-),
\eeq
which means that we can get $\Gamma(W^-\to V\ell^-\bar{\nu}_\ell)/\Gamma_0$ using the experimental data of $\Gamma(V\to e^+e^-)$ given by Particle Data Group \cite{PDG22} directly. Numerical results have been listed in Table 1, and the errors of the predictions in the fifth column are due to the uncertainties of the measured widths of  $\Gamma(V\to e^+e^-)$ only. To transform them into the branching fractions of $W\to V \ell\bar{\nu}_\ell$, one may use the experimental data of ${\cal B}(W\to \ell \bar{\nu}_\ell)$, which can be found in Ref. \cite{PDG22} that
\beq\label{brW-ell-nu}
{\cal B}(W^-\to e^-\bar{\nu}_e)=(10.71\pm 0.16)\%,\;\;\;\;\;\; {\cal B}(W^-\to \mu^-\bar{\nu}_\mu)=(10.63\pm 0.15)\%.
\eeq
For our numerical analysis, we take
\beq\label{brW-ell-nu-avg}
{\cal B}(W^-\to\ell^-\bar{\nu}_\ell)=(10.67\pm 0.16)\%
\eeq
by simply averaging over the electron and muon modes. Thus, it is straightforward to obtain the branching fractions of rare $W$-boson decays into a vector meson and lepton pair, for $\ell=e$ or $\mu$, which read
\beqn\label{brW-rho-ell-nu}
{\cal B}(W^-\to \rho\ell^-\bar{\nu}_\ell)=(5.64\pm 0.10)\times 10^{-6},\\
\label{brW-omega-ell-nu}
{\cal B}(W^-\to \omega\ell^-\bar{\nu}_\ell)=(4.74\pm 0.17)\times 10^{-7},\\
\label{brW-phi-ell-nu}
{\cal B}(W^-\to \phi\ell^-\bar{\nu}_\ell)=(6.60\pm 0.23)\times 10^{-7},\\
\label{brW-jpsi-ell-nu}
{\cal B}(W^-\to J/\Psi\ell^-\bar{\nu}_\ell)=(4.24\pm 0.10)\times 10^{-7}.
\eeqn
Here the quoted errors of our theoretical results show the uncertainties from the experimental values of $\Gamma(V\to e^+e^-)$ in the third column of Table 1, and also ${\cal B}(W^-\to \ell^-\bar{\nu}_\ell)$ in eq. (\ref{brW-ell-nu-avg}).

It is found that branching ratios of $W\to V \ell\bar{\nu}_\ell$ decays obtained in the present work are quite larger than those of the hadronic radiative decays $W^\pm\to M^\pm\gamma$ ($M$ is a pseudoscalar or vector meson such as $\pi$, $K$, $\rho$, $K^*$, and $D_s$ etc), which are maximally around $10^{-8}$ or even smaller, predicted by the authors of Ref. \cite{GKN15}. Naively, one may expect that $\Gamma(W\to V\ell\bar{\nu}_\ell)$ should be smaller than $\Gamma(W^\pm\to M^\pm\gamma)$ since the former rate is suppressed by a power of $\alpha_{\rm em}$ compared to the latter rate. However, careful observation can tell us this expectation is not correct. As given in Ref. \cite{GKN15}, we know
 \beq\label{rate-WMGamma} \Gamma(W^\pm\to M^\pm\gamma)\sim \frac{\alpha_{\rm em}f_M^2}{192 m_W}. \eeq
Comparing with eq. (\ref{rate1}), one will find a relevant factor $m_W^2/m_V^2$ in the formula of $\Gamma(W^-\to V\ell^-\bar{\nu}_\ell)$, which could significantly counteract the suppression of $\alpha_{\rm em}$ if the mass of vector meson is very small relative to the $W$ mass. Obviously, the appearance of this factor is actually due to the virtual photon propagator of $\gamma^*\to V$ transition in Figure 1.

Similar situation also occurs in rare $Z$-boson decays.  In particular, it has been shown in Ref. \cite{Fleming93-94} that the dominant contribution to $Z\to V\ell^+\ell^-$ comes from $Z\to \gamma^*\ell^+\ell^-$ with the subsequent transition $\gamma^*\to V$, since, in comparison, the radiative decays $Z\to V\gamma$ are quite suppressed. One can thus neglect the contribution from $Z\to V\gamma^*\to V\ell^+\ell^-$ although it is of the same order of $\alpha_{\rm em}$ as the dominant part.

Analogous to $Z\to V\gamma^*\to V\ell^+\ell^-$, the rare charged weak gauge boson decays considered in the present paper could happen through $W\to V W^*\to V \ell\bar{\nu}_\ell$. The Feynman diagram has been displayed in Figure 2, and we take $V=J/\Psi$ as an explicit example. As a good approximation for the leading order calculation, the momenta of the quark ($c$) and anti-quark ($\bar{c}$) are taken to be one half of $J/\Psi$ momentum $q$, so the strange quark propagator in this diagram is proportional to $1/(k+\frac{q}{2})^2$, which is of order $1/m_W^2$. By contrast, the virtual photon propagator in the diagrams of Figure 1 is of order $1/m^2_{J/\Psi}$. This means that the contribution from Figure 2, relative to that from Figure 1, is strongly suppressed, which can be safely neglected.

\begin{figure}[t]
\begin{center}
\includegraphics[width=7cm,height=3cm]{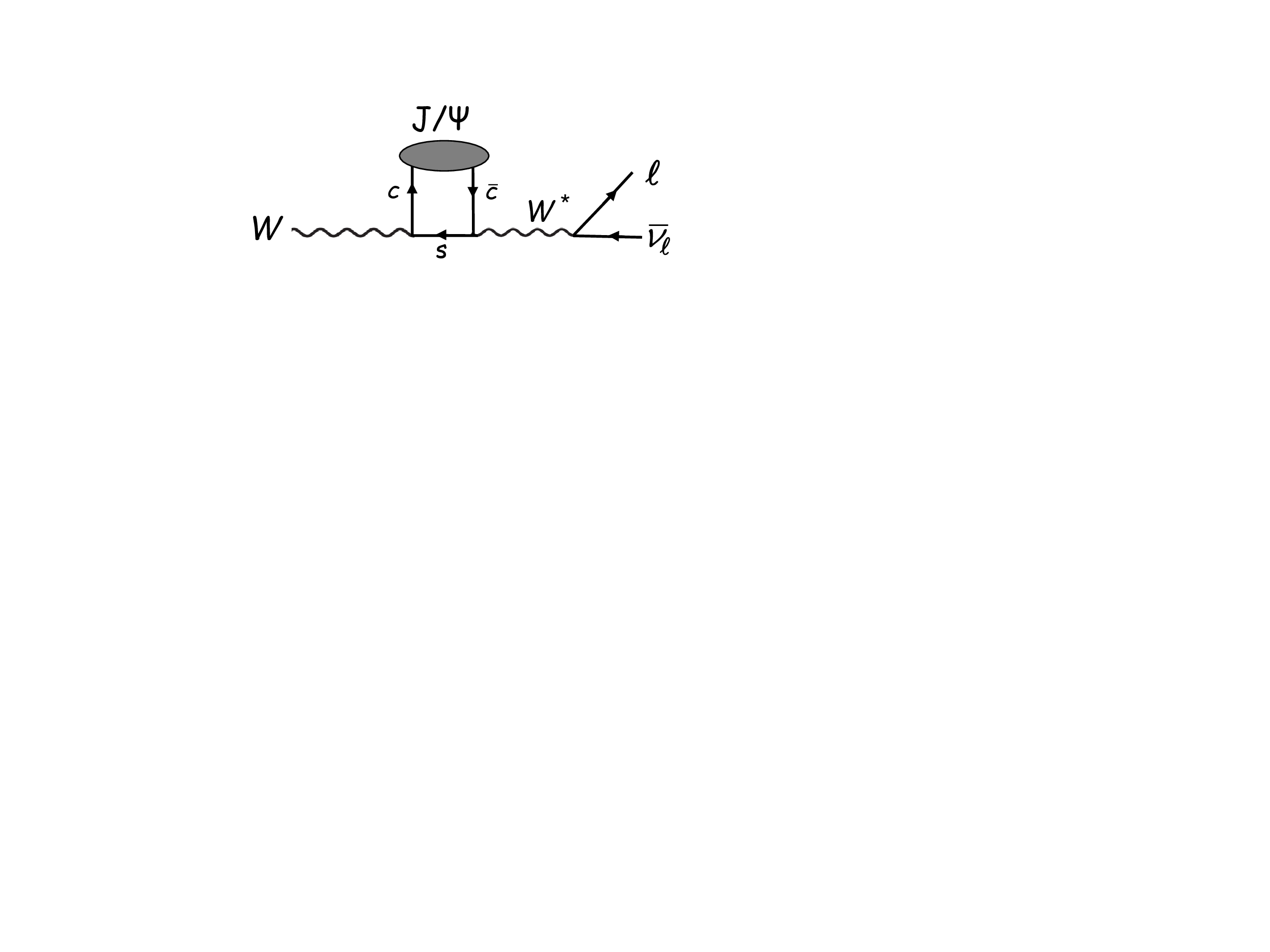}
\end{center}
\caption{The Feynman diagram contributing to $W\to J/\Psi \ell \bar{\nu}_\ell$ decays via $W\to J/\Psi W^*$ transition.}\label{figure2}
\end{figure}

Furthermore, recall that the differential decay rate of $W^-\to V\ell^-\bar{\nu}_\ell$ has been given in eq. (\ref{rate1}). Now one can rewrite
\beq\label{kinematical-variables-in-Lab}
s_V=1+r_V^2-2 E_V/m_W,\;\;\;\;\;\; s_\ell=1-2 E_\ell/m_W,
\eeq
where $E_V$ is the vector meson energy and $E_\ell$ is the lepton energy in the rest frame of $W$ boson. In terms of $E_V$ and $E_\ell$, we have
\beq\label{rate4}
\frac{d\Gamma}{dE_V dE_\ell}=\frac{\alpha_{\rm em}^2 Q_V^2 g^2 f_V^2}{96\pi m_W^3 r_V^2}I_V.
\eeq
Thus the energy spectrum of the rare decays can be obtained by integrating over $E_V$ or $E_\ell$. The normalized energy distributions of $W^-\to J/\Psi \ell^-\bar{\nu}_\ell$ with respect to $E_J$ and $E_\ell$ have been plotted in Figure 3, respectively. The peak of the distribution is corresponding to the small $J/\psi$ energy or large lepton energy region. Since we have neglected the lepton mass in the calculation, the spectrum in left plot does not go to zero for $E_J$ at $\sim m_W/2$. We are not going to display the plots for the differential rate of $W^-\to V \ell^-\bar{\nu}_\ell$ decays when $V$ is the light vector meson ($\rho$, $\omega$, and $\phi$) because it is believed that one will achieve the similar behavior as above.

\begin{figure}[t]
\begin{center}
\includegraphics[width=7.5cm,height=5cm]{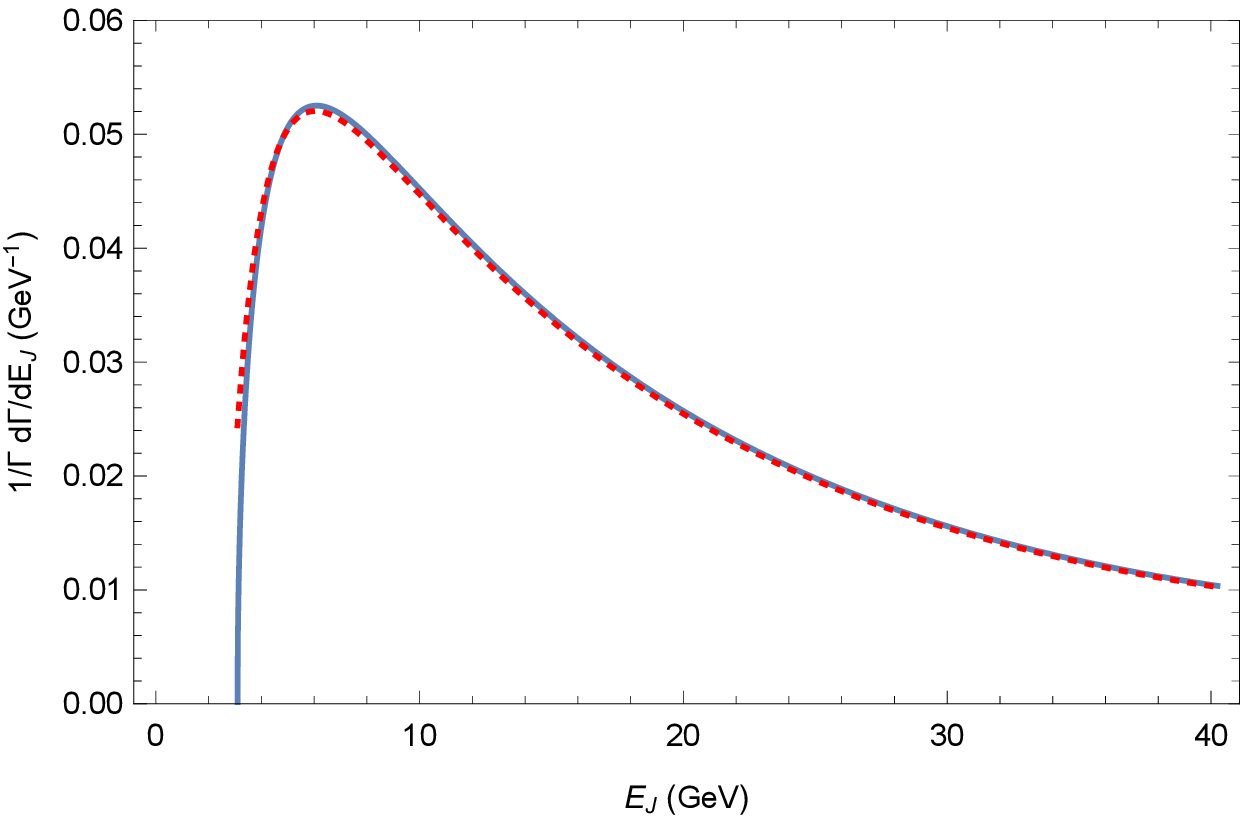}\hspace{0.25in}
\includegraphics[width=7.5cm,height=5cm]{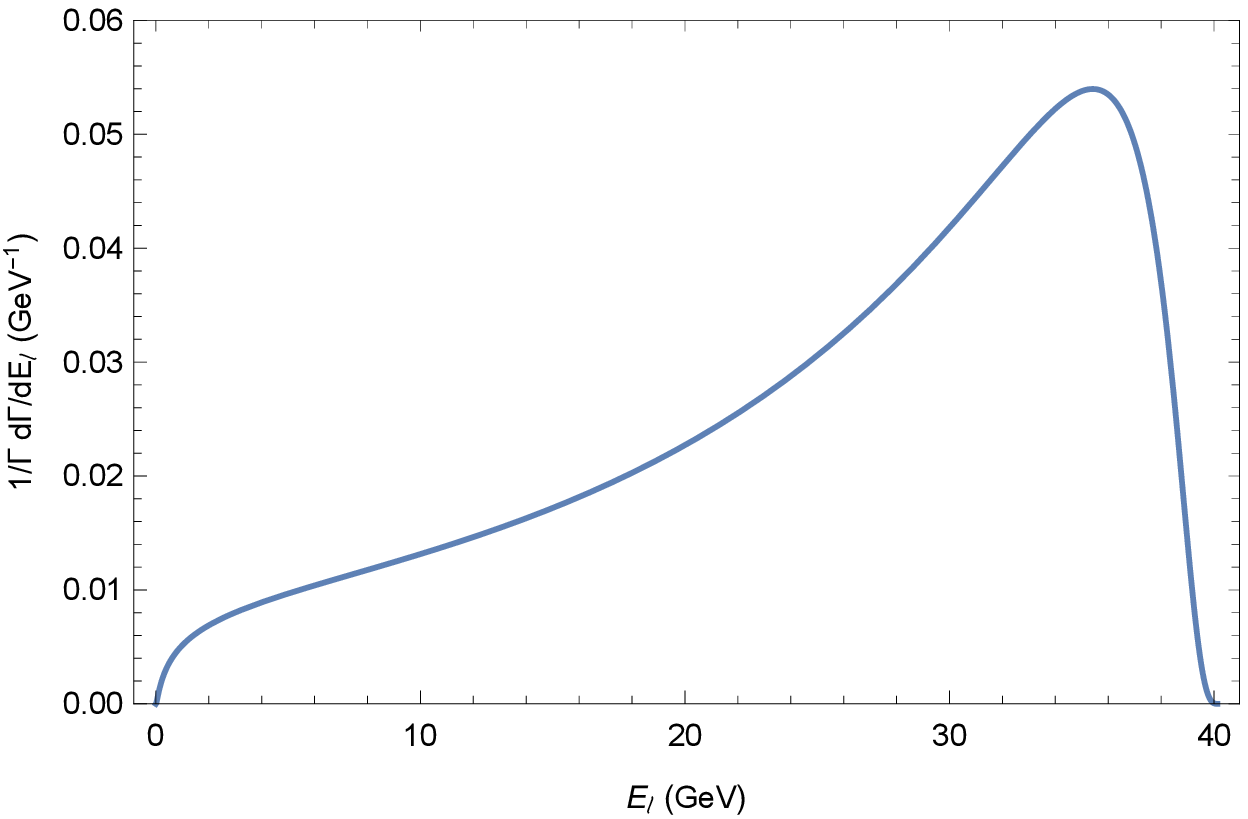}
\end{center}\caption{The normalized energy spectrum of  $W\to J/\Psi \ell^- \bar{\nu}_\ell$ decays with respect to $J/\Psi$ energy $E_J$ (left plot), and with respect to the lepton energy $E_\ell$ (right plot). The red-dashed line in the left plot is from eq. (\ref{fragrate}).}\label{figure3}
\end{figure}

It has been pointed out in Ref. \cite{Fleming93-94} that the large rate of $Z\to J/\Psi \ell^+\ell^-$ decays can be explained by a fragmentation contribution which is not suppressed by a factor of $m_{J/\Psi}^2/m_Z^2$. As shown in the second paper of Ref. \cite{Fleming93-94} explicitly, in the fragmentation limit $m_Z\to \infty$ with $E_J/m_Z$ fixed, the rate for the processes could be factorized into electromagnetic decay rates and universal fragmentation function, and particularly, the differential decay rate $d\Gamma/d E_J$ obtained in this way agrees with the full calculation in the fragmentation limit. Similar study can be done in the present work by borrowing this approach. After integrating over $E_\ell$ in eq. (\ref{rate4}) and taking the limit $m_W\to \infty$ with $E_V/m_W$ fixed, we obtain
\beq\label{fragrate}\frac{d\Gamma}{d E_V}=\frac{2 \alpha_{\rm em}^2 Q_V^2 g^2 f_V^2}{m_V^2}\frac{\Gamma(W^-\to \ell^-\bar{\nu}_\ell)}{m_W}\left[\frac{(x-1)^2+1}{x}\log{\frac{x^2}{r_V^2}}-\frac{2}{x}+2-\frac{8}{3}x\right],
\eeq
where $x=2 E_V/m_W$. It is found that the distribution for $J/\Psi$ energy $E_J$, which has been displayed in Figure 3, is very closed to the plot from the full calculation. Certainly, a systematical and detailed analysis of electromagnetic fragmentation in $W^-\to J/\Psi \ell^-\bar{\nu}_\ell$ decays would be an interesting topic. This is left for a future separate publication.

To summarize, we have presented the analysis of exclusive rare $W$-boson decays into a vector meson and lepton pair. In the SM, the leading order contributions to these processes come from $W\to \gamma^*\ell\bar{\nu}_\ell$, followed by $\gamma^*\to V$. Using the measured widths of $\Gamma(V\to e^+e^-)$ given in \cite{PDG22}, we have determined the branching fractions of $W^-\to V\ell^-\bar{\nu}_\ell$ for $V=\rho$, $\omega$, $\phi$, and $J/\Psi$, respectively, as shown in eqs. (\ref{brW-rho-ell-nu}) -- (\ref{brW-jpsi-ell-nu}). It is surprising that branching fractions of these three-body decays, although they are suppressed by a power of $\alpha_{\rm em}$, are quite larger than those of two-body hadronic radiative decays $W^\pm\to M^\pm\gamma$, which have been predicted by the authors of Ref. \cite{GKN15} already. Furthermore, note that the $\gamma WW$ vertex, as shown in Figure 1(b), is involved in the transition, thus both experimental and theoretical investigations of $W\to V\ell\bar{\nu}_\ell$ decays may be also helpful to test triple gauge couplings.

Our experimentalists have been trying to search for exclusive rare $W$-boson processes containing hadronic final states. Unfortunately, so far no such decays have been observed.
Theoretical predictions on branching fractions of $W\to V\ell\bar{\nu}_\ell$ in the present paper are around $10^{-6}\sim 10^{-7}$. Experimentally, the heavy quarkonium $J/\psi$ is in general reconstructed via leptonic decays with their rates: ${\cal B}(J/\Psi\to \ell^+\ell^-)=(5.971\pm 0.032)\%$ \cite{PDG22}; while for light vector mesons, $\rho$ decays almost exclusively to $\pi^+\pi^-$, $\omega$ and $\phi$ have a large rate into $\pi^+\pi^-\pi^-$ and $K^+ K^-$, respectively, in the event construction. Therefore,
our analysis seems to indicate that these exclusive rare $W$ decay modes could be the promising candidates in future experimental machines, for instance, in the high-luminosity LHC, where large amount of $W$ bosons about ${\cal O}(10^{11})$ events will be produced. We eagerly await some dedicated searches for such decays at these facilities.

\section*{Acknowledgments}
This work was supported in part by the National Natural Science Foundation of China under Grants No. 11575175, No. 12047502, and No. 12247103, and by National Research and Development Program of China under Contract No. 2020YFA0406400.

\appendix
\newcounter{pla}
\renewcommand{\thesection}{\Alph{pla}}
\renewcommand{\theequation}{\Alph{pla}\arabic{equation}}
\setcounter{pla}{1}
\setcounter{equation}{0}

\section*{Appendix: Explicit expression of $I_V$ }

 After squaring the $W^-\to V \ell^-\bar{\nu}_\ell$ decay amplitude and summing/averaging spins of all particles, we get the differential decay rate  of eq. (\ref{rate1}) as
 $$\frac{d\Gamma}{d s_V d s_\ell}=\frac{\alpha_{\rm em}^2 Q_V^2 g^2 f_V^2}{384\pi m_W r_V^2}I_V.$$
The full expression of $I_V$, in terms of dot product of the relevant four-momenta, can be given by
\beq\label{I}
I_V
=I_1+I_2+I_3,
\eeq
where
\beq\label{I1}
I_1=\frac{8}{(q^2+2 k_1\cdot q)^2}\left[(2 k_1\cdot q k_2 \cdot q - q^2 k_1\cdot k_2)+\frac{2p\cdot k_2}{m_W^2}(2 k_1\cdot q p\cdot q-q^2 k_1\cdot p)\right],
\eeq
\beqn\label{I2}
I_2=\frac{8}{(q^2+2 k_1\cdot q)(q^2+2 k\cdot q)}\left( 2 k_1\cdot k_2(4k_1\cdot k_2+2 k\cdot q+4 q^2)-4 k_1\cdot q k_2\cdot q\right.\nonumber\\
    -\frac{1}{m_W^2}[(2k_1\cdot q+4 k\cdot q+3 q^2)(2 k_1\cdot q k_2\cdot q-q^2 k_1\cdot k_2)\nonumber \\
    \left. -4 p\cdot k_2 (q^2+2 k_2\cdot q)k_1\cdot k_2+4 p\cdot q (q^2+2 k_1\cdot q)k_1\cdot k_2]\right),
\eeqn
and
\beqn\label{I3}
I_3=\frac{8}{(q^2+2 k\cdot q)^2}\left[12k_1\cdot q k_2\cdot q-((2k+q)^2+10 q^2)k_1\cdot k_2 \right.\nonumber\\
\left. -\frac{1}{2m_W^2}(2k+q)^2(2 k_1\cdot q k_2\cdot q-q^2 k_1\cdot k_2)+\frac{4(p\cdot q)^2}{m_W^2}\right].
\eeqn
On the other hand, one can easily get
\beqns
k_1\cdot k_2=\frac{m_W^2}{2}s_V,\;\;\;\; p\cdot q=\frac{m_W^2}{2}(1+r_V^2-s_V),\\
k_1\cdot q=\frac{m_W^2}{2}(1-s_V-s_\ell), \;\;\;\;k_2\cdot q=\frac{m_W^2}{2}(s_\ell-r_V^2),\\
k_1\cdot p=\frac{m_W^2}{2}(1-s_\ell),\;\;\;\; k_2\cdot p=\frac{m_W^2}{2}(s_V+s_\ell-r_V^2).
\eeqns
For on-shell initial and final states particles, we could take $p^2=m_W^2$, $q^2=m_V^2$, and $k_1^2=k_2^2=0$ (lepton masses are set to be zero already).
This shows that $I_V$ can be in terms of the kinematical variables $s_V$ and $s_\ell$ completely.

\end{document}